\documentclass[aip, apl, graphicx, numerical, reprint, twocolumn,amsmath, amssymb]{revtex4-1}

\usepackage{graphicx}
\usepackage{bm}
\usepackage{natbib}
\usepackage{amssymb}
\usepackage{color}

\begin{document}

\title{\textcolor{blue} {First principles calculation of polarization induced interfacial charges in GaN/AlN heterostructures}} 

\author{Rohan Mishra}
\affiliation{Department of Materials Science and Engineering, The Ohio State University, Columbus, Ohio 43210}
\author{Oscar D. Restrepo}
\affiliation{Department of Materials Science and Engineering, The Ohio State University, Columbus, Ohio 43210}
\author{Siddharth Rajan}
\affiliation{Department of Electrical and Computer Engineering, The Ohio State University, Columbus, Ohio 43210}
\affiliation{Department of Materials Science and Engineering, The Ohio State University, Columbus, Ohio 43210}
\author{Wolfgang Windl}
\affiliation{Department of Materials Science and Engineering, The Ohio State University, Columbus, Ohio 43210}

\date{\today}

\begin{abstract}
We propose a new method to calculate polarization induced interfacial charges in semiconductor heterostructures using classical electrostatics applied to real-space band diagrams from first principles calculations and apply it to
GaN/AlN heterostructures with ultrathin AlN layers (4-6 monolayers). We show that the calculated electric fields and interfacial charges are independent of the exchange-correlation functionals used (LDA and HSE06).  
We also find the calculated interfacial charge of $(6.8\pm0.4)\times10^{13}$ cm$^{-2}$ to be in excellent agreement with experiments and  the value of $6.58\times10^{13}$ cm$^{-2}$ calculated from bulk polarization constants, validating the use of bulk constants even for very thin films.
\end{abstract}

\pacs{}

\maketitle 
Gallium nitride (GaN) and aluminum nitride (AlN) are binary III/V semiconductors with wide direct band gaps of $3.4$ and $6.2$ eV, respectively, making them ideal candidates for applications in optoelectronic, high-power and high-frequency devices. In addition, group III-nitride heterostructures have strong spontaneous polarization ($P_{\rm SP}$) due to their lower-symmetry wurtzite structure and piezoelectric polarization ($P_{\rm PE}$) due to their large lattice mismatch. This large macroscopic polarization leads to a high density of charge at the heterointerface. Polarization engineering has been successful to induce high sheet density electron channels without the disadvantages of impurity doping, enabling new routes for device design. Scaling of high electron mobility transistors 
for higher frequency applications has been shown to require the use of ultrathin AlGaN and AlN cap layers with thicknesses between 0.5 nm and 20 nm.\cite{Keller_2002,Cao_2007}
 
To date, polarization in heterostructures has been modeled using bulk constants for the calculation of the interface charge density 
and validated only for one case of thin GaN and AlN layers with equal thickness.\cite{Bernardini_PRB_57} 
However, the validity of this approximation 
for arbitrary 
\textquotedblleft non-bulk\textquotedblright layer thicknesses is still an open question.\cite{Bernardini_PRB_57,Nardelli_1997}
 
In this letter, we propose a novel method to calculate interface charges, where classical electrostatics are applied to real-space band diagrams obtained from density-functional theory (DFT) calculations and use it to calculate the interface charges in GaN/AlN heterojunctions with ultrathin AlN layers (4-6 monolayers (MLs) thick). We show that this method is more convenient and consistent for calculating interface charges than the commonly used Bader analysis.\cite{Bader_1990} Our results are finally used to examine the validity of using bulk constants in the limit of small film thicknesses.

Within the traditional bulk approximation,\cite{Bernardini_PRL_79,*Bernardini_PRB_63, *Bernardini_PSS_216, Bernardini_PRB_56} the polarization induced interface charges in GaN/AlN thin films are calculated from DFT-derived polarization coefficients\cite{King-smith_PRB_47}  and elastic constants. There, the charge density $\sigma_{\rm SP}$ induced by spontaneous polarization $P_{\rm SP}$ is 
\begin{eqnarray}
\sigma_{\rm SP}={P_{\rm SP}^{\rm AlN}}-{P_{\rm SP}^{\rm GaN}},
\label{eq1}
\end{eqnarray}

Using the values of ${P_{\rm SP}^{\rm AlN}}= -0.081$ cm$^{-2}$ and ${P_{\rm SP}^{\rm GaN}} = -0.029$ cm$^{-2}$ from Ref.~\onlinecite{Bernardini_PRB_56}, we get $\sigma_{\rm SP} = 3.25\times10^{13}$ cm$^{-2}$.  Similarly, the interface charge density $\sigma_{\rm PE}$ induced by the piezoelectric polarization $P_{\rm PE}$ is
\begin{eqnarray}
\sigma_{\rm PE}=e_{33}\epsilon_3+2e_{31}\epsilon_1,
\label{eq2}
\end{eqnarray}
with piezoelectric constants $e_{33}$ and $e_{31}$, in-plane and out-of-plane strains $\epsilon_1=(a_{\rm GaN}-a_{\rm AlN})/a_{\rm AlN}$ and $\epsilon_3=-2\epsilon_1c_{13}/c_{33}$, and AlN elastic constants $c_{13}$ and $c_{33}$. With Eqs.~(\ref{eq1}) and (\ref{eq2}), the total charge density $\sigma_{\rm SP} + \sigma_{\rm PE}$ is then independent of the layer thicknesses.

For our DFT calculations, we used the 
VASP code\cite{Kresse_PRB_47,*Kresse_PRB_49} with projector augmented wave (PAW) potentials\cite{Blochl_PRB_50} within the local density approximation (LDA)\cite{Ceperley_APL_45} 
as had been used in previous works.\cite{Bernardini_PRB_57,Nardelli_1997}
 Additionally, as LDA underestimates band-gaps of solids by $30\%-100\%$,\cite{Wang_PRL_51} we repeat some band structure calculations with a more accurate (but computationally expensive) 
modified 
HSE06 (mHSE06) hybrid exchange-correlation (XC) functional \cite{Heyd_JCP_118,*Heyd_JCP_124,*Paier_JCP_124} for PBE\cite{PBE,*PBE_errata}-relaxed structures. We use 
an optimized Hartree-Fock fraction of $\alpha = 0.35$ 
to get good agreement with the experimental bandgaps of \textit{both} GaN and AlN. 
A plane-wave cutoff energy of 400 (500) eV was used for LDA (mHSE06) calculations, along with $\Gamma$-centered  Monkhorst-Pack  meshes\cite{Monkhorst_PRB_13}  with divisions $N_i$ such that the product of $N_i$ with the corresponding lattice constant $a_i$ was 25, 60, and 12~{{\AA}} for LDA relaxations, LDA band structures, and mHSE06 calculations, respectively. The Ga $d$-orbital electrons were treated as core (valence) electrons for LDA (mHSE06) calculations. 

Periodic $(110)\times(\overline{1}10)\times(00n)$, $n=4,...14$, superlattices along the [001] direction of the wurtzite-structure were used to simulate the different heterostructures. Ultrathin AlN layers consisted of 4 to 6 MLs in combination with 4 to 24 MLs of GaN. To simulate thin films on GaN substrate, the in-plane lattice constants of the supercells were kept fixed to match that of relaxed GaN while allowing the out-of-plane constants to relax.

The band diagram along the $z$-axis in real space was obtained by averaging over the atom-projected DOS of all atoms in the layers between $z$ and $z+
2.4$ {\AA}. The exact band edges of the projected DOS were validated with the help of band structure calculations in reciprocal space. 

The LDA optimized lattice parameters for GaN ($a=3.17$ {\AA}, $c=5.16$ {\AA}) and AlN ($a=3.09$ {\AA}, $c=4.95$ {\AA}) agree within less than 1\% with the experimental values, $a=3.16$ {\AA}, $c=5.13$ {\AA}, for GaN\cite{Lagerstedt_PRB_19} and $a=3.11$ {\AA}, $c=4.98$ {{\AA}} for AlN.\cite{Iwama_JCG_56} The bulk band gaps for GaN and AlN were found to be 2.13 eV and 4.53 eV, respectively, compared to experimental values of 3.4 eV and 6.2 eV. The mHSE06 band gaps for PBE relaxed lattice constants of $a=3.22$ {\AA}, $c=5.24$ {\AA}  for GaN and $a=3.13$ {\AA}, $c=5.01$ {\AA} for AlN are 3.61 eV and 6.04 eV, respectively. Our LDA values for $c_{13}$ and $c_{33}$ of AlN are 113 GPa and 371 GPa, respectively, in good agreement with previous calculations\cite{Wright_JAP_1997,*Shimada_JJAP_2006} and experimental values of $99\pm 4$ GPa and $389\pm 10$ GPa.\cite{McNeil_JACerS_76} 
Combining them with the $P_{\rm PE}$ constants from Ref.~\onlinecite{Bernardini_PRB_56}, we find $\sigma_{\rm PE} = 3.33\times 10^{13}$ cm$^{-2}$. Adding $\sigma_{\rm SP}$ and $\sigma_{\rm PE}$, we find the total interface electron density to be $6.58\times 10^{13}$ cm$^{-2}$, independently of the layer thicknesses.
\begin{figure}
\includegraphics[width=1.0 \columnwidth, angle=-90, scale=0.9]{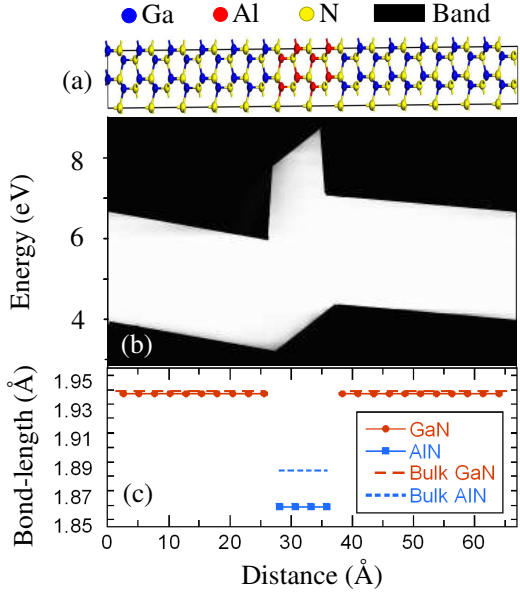} 
\caption{(Color online) Relaxed (22 ML GaN)/(4 ML AlN) heterostructure (a), its layer projected density of states (b), and its variation of bond lengths along the [001] direction (c).}
\label{fig:Figure1} 
\end{figure}
In the band diagram of a heterostructure with 4 MLs of AlN and 22 MLs of GaN (Fig.~\ref{fig:Figure1}), we find that band bending is affected by both $P_{\rm SP}$ and $P_{\rm PE}$ and is a strong function of bond lengths. The electric field $\xi$ obtained from the average slope of the bands $dE/dz$ divided by the electron charge $e$ is found to be $-1.87\times 10^6$ V/cm and $1.11\times 10^7$ V/cm in the GaN and AlN layers, respectively. We calculate the interface electron density $\sigma_{\rm {int}}$ from applying the integral form of Gauss' law\cite{Griffiths} to a volume including the interface, resulting in
\begin{eqnarray}
\sigma_{\rm int}=\varepsilon_{\rm GaN}\xi_{\rm GaN}-\varepsilon_{\rm AlN}\xi_{\rm AlN}.
\label{eq3}
\end{eqnarray}
Using experimental permittivities $\varepsilon_{\rm GaN}=7.88\times 10^{-11}$~F/m and $\varepsilon_{\rm AlN}=8.01\times 10^{-11}$~F/m, we find $\sigma_{\rm {int}}({\rm GaN/AlN}) = -\sigma_{\rm int}({\rm AlN/GaN}) =(6.6\pm 0.3)\times 10^{13}$ cm$^{-2}$. The error bars are due to the uncertainty in finding the slope of the bands accurately as a result of the very low DOS present at the band edges. From the different supercells used, we find $\sigma_{\rm {int}}$ to be independent of the thickness of the GaN layer with an average value of $(6.7\pm 0.3)\times 10^{13}$ cm$^{-2}$ for supercells with 4 AlN MLs and $(6.9\pm 0.3)\times 10^{13}$ cm$^{-2}$ for the ones with 6 AlN MLs. Since the two values agree within their errorbars, we take an average value of    $(6.8\pm 0.4)\times 10^{13}$ cm$^{-2}$ as the charge density at the  GaN/AlN interface, which is at the upper end of the experimental range of (1 to 6)$\times 10^{13}$ cm$^{-2}$.\cite{Keller_2002,Cao_2007}
 
In order to examine the dependence of our results on the value of the band-gap, we repeat the interface charge density calculations for (4 ML AlN/12 ML GaN) and (6 ML AlN)/(8 ML GaN) heterostructures with the mHSE06 functional.
We find mHSE06 charge densities of  $(6.7\pm 0.3)\times 10^{13}$ cm$^{-2}$ and $(6.6\pm 0.3)\times 10^{13}$ cm$^{-2}$, respectively, as compared to LDA values of $(6.4\pm 0.3)\times 10^{13}$ cm$^{-2}$ and $(6.8\pm 0.3)\times 10^{13}$ cm$^{-2}$. Thus, the influence of the XC functional used on the slope of the bands in these heterostructures is found to be negligible within the computational error bars, allowing the calculation of interface charges even for relatively large systems (with 100s of atoms) using the much faster LDA functional.

In order to corroborate our calculations, we determine the variation of the electric field with GaN layer with an alternative approach.
The periodic boundary conditions in our calculations along with the conservation of charges require that the integrated electric field, or overall electrostatic potential, is zero. Thus, Eq.~(\ref{eq3}) requires that the charges at the GaN/AlN and AlN/GaN interfaces are of equal size, but opposite sign. Since the average slopes of the band edges in the GaN and AlN layers away from the interface are constant, this condition then leads to	 
\begin{eqnarray}
\xi_{\rm GaN}t_{\rm GaN}=\xi_{\rm AlN}t_{\rm AlN},
\label{eq4}
\end{eqnarray}
where $t_{\rm GaN}$ and $t_{\rm AlN}$ are the thicknesses of the GaN and AlN layers, respectively, measured from one of the interfaces. Combining Eqs.~(\ref{eq3}) and (\ref{eq4}), we find 
 \begin{eqnarray}
\xi_{\rm GaN}={\sigma_{\rm int}}/\left[{\varepsilon_{\rm GaN}(1+{t_{\rm GaN}}/{t_{\rm AlN}})}\right].
\label{eq5}
\end{eqnarray}
\begin{figure}
\includegraphics[width=1.0 \columnwidth, angle=0, scale=0.8]{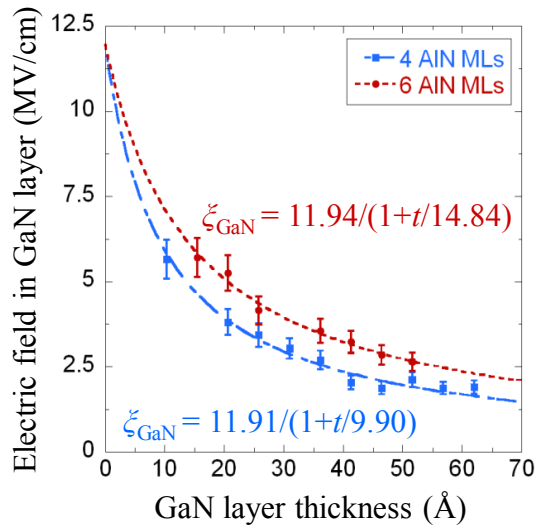} 
\caption{(Color online) Variation of electric field in the GaN layer with changing GaN layer thickness for two fixed AlN interlayer thicknesses of 4 (blue squares) and 6 (red circles) MLs from LDA calculations. The lines are fits using Eq.~(5).}\label{fig:Figure2} 
\end{figure}
Figure \ref{fig:Figure2} shows a plot of the electric field in the GaN layer vs.\ $t_{\rm GaN}$ and a corresponding fit of Eq.~(\ref{eq5}) for heterostructures with 4 and 6 AlN MLs. The error bars are due to the uncertainty in delineating the band edges. With $t_{\rm AlN}$ fixed, the only adjustable parameter in Eq.~(\ref{eq5}) is  $\sigma_{\rm int}$. Equation (\ref{eq5}) fits the calculated electric field values well with a least-squares value of  $\sigma_{\rm int}=5.85\times 10^{13}$ cm$^{-2}$. By extrapolating the curve to $t_{\rm GaN}=0$, we get the limiting value of the electric field in the GaN layer to be   $1.19\times 10^7$V/cm (with a negligible difference of $0.03$ MV/cm between the supercells with 6 and 4 MLs of AlN). Both values are in reasonable agreement with the values determined above from Eq.~(\ref{eq3}),$(6.8\pm 0.4)\times 10^{13}$ cm$^{-2}$ and $(1.38\pm 0.08)\times 10^7$ V/cm, respectively, thus demonstrating the consistency of our approach.

As a further check, we also calculate the interface charge density directly from the DFT charge distribution at the interface.\cite{Bernardini_PRB_57} For that, we use the Bader approach,\cite{Henkelman_CMS_36} where the number of electrons on each atom is determined by integrating the charge density over the volume enclosed by the atom's zero flux surface. Charges on atoms at defects can then be determined relative to the charge on perfect bulk atoms, which is $1.85e$ and $1.06e$ for N atoms in bulk AlN and GaN, respectively. Reference charges for interfacial N with 3 Al and 1 Ga bonds and vice versa are calculated from a structure with alternating GaN and AlN MLs with values of $1.61e$ and $1.19e$ respectively (due to the lack of any other conventional definition). This treatment gives the average interface charge density to be $5.95\times 10^{13}$ cm$^{-2}$ and $6.50\times 10^{13}$ cm$^{-2}$ for the heterostructures with 4 MLs and 6 MLs of AlN (with different monolayers of GaN), respectively which is close to the values that we got from the electrostatics treatment. Again, since there is not a significant change in the charge density with 4 and 6 MLs of AlN, we take an average value of $6.23\times 10^{13}$ cm$^{-2}$ as the interface charge density in GaN/AlN heterostructures.
 
One of the central questions of this letter concerns the validity of using bulk polarization constants to calculate interface charges in the limit of ultrathin layers. Since the value of interface charge density for the GaN/AlN heterostructures with ultrathin AlN layers obtained with our approach,$(6.8\pm 0.4)\times 10^{13}$ cm$^{-2}$, agrees well with the value of $6.58\times 10^{13}$ cm$^{-2}$ from bulk polarization constants, our results suggest that the use of bulk polarization constants to calculate interface charges is a valid approach even for heterostructures with very few MLs.
 
In conclusion, we have calculated the interfacial polarization charge density in GaN-AlN heterostructures from a combination of DFT with  electrostatics and find $\sigma_{\rm int}=(6.8\pm 0.4)\times 10^{13}$ cm$^{-2}$, at the upper end of the experimental range of (1 to 6)$\times 10^{13}$ cm$^{-2}$.\cite{Keller_2002,Cao_2007} This value is in the same range as the charge densities determined from the dependence of the electric field in GaN on the layer thickness,  $5.85\times 10^{13}$ cm$^{-2}$, and from a Bader analysis, $6.23\times 10^{13}$ cm$^{-2}$. The latter however has uncertainties about interfacial reference charges and number of layers to include into the charge density calculation. In comparison, we find that the combination of ab-initio real-space band structures and electrostatics eliminates these uncertainties and is applicable to a wide range of systems beyond the AlN/GaN system studied. Our value agrees well with the interface charge density calculated using bulk constants of $6.58\times 10^{13}$ cm$^{-2}$, thus validating the use of the former in heterostructures where layer thicknesses approach \textquotedblleft non-bulk\textquotedblright dimensions. Additionally, this method could also be used to calculate charges at more complex interfaces, such as, among others, in digital superlattices and tunnel barrier systems.

This work was supported by the
Institute for Materials Research and the NSF 
MRSEC at OSU (Award No. DMR-0820414) . Computations were performed at the Ohio Supercomputer Center (Grant No. PAS0072). We thank Digbijoy Nath for 
discussions.
\bibliography{References_ww}
\end{document}